\documentclass[aps,prb,twocolumn,floatfix,amsmath,amssymb,superscriptaddress]{revtex4}
\usepackage{graphicx}
\usepackage{color}
\usepackage{dcolumn}
\usepackage{bm}
\usepackage{epsfig,psfrag,amsmath,amssymb,float}
\input{epsf}

\begin{document}

\title{Pairing in graphene: A quantum Monte Carlo study}
\author{Tianxing Ma}
\affiliation{Department of Physics, Beijing Normal University,
Beijing 100875, China}
\affiliation{Beijing Computational Science Research Center,
Beijing 100084, China}
\author{Zhongbing Huang}
\email{huangzb@hubu.edu.cn}
\affiliation{Faculty of Physics and Electronic Technology, Hubei University,
Wuhan 430062, China}
\affiliation{Beijing Computational Science Research Center,
Beijing 100084, China}
\author{Feiming Hu}
\affiliation{Department of Physics and ITP, The
Chinese University of Hong Kong, Hong Kong}
\author{Hai-Qing Lin}
\email{hqlin@phy.cuhk.edu.hk}
\affiliation{Beijing Computational Science Research Center,
Beijing 100084, China}
\affiliation{Department of Physics and ITP, The
Chinese University of Hong Kong, Hong Kong}
\date{\today}

\begin{abstract}
To address the issue of electron correlation driven superconductivity
in graphene, we perform a systematic quantum Monte Carlo study of the
pairing correlation in the $t-U-V$ Hubbard model on a honeycomb lattice.
For $V=0$ and close to half filling, we find that pairing with $d+id$
($d_{x^{2}-y^{2}}+id^{\prime}_{xy}$ in its specific form) symmetry
dominates pairing with extended-$s$ symmetry. However, as the system
size or the on-site Coulomb interaction increases, the long-range part
of the $d+id$ pairing correlation decreases and tends to vanish in the
thermodynamic limit. An inclusion of nearest-neighbor interaction $V$,
either repulsive or attractive, has a small effect on the extended-$s$
pairing correlation, but strongly suppresses the $d+id$ pairing correlation.
\end{abstract}

\maketitle
Recently, graphene has attracted the attention of experimentalists and
theorists\cite{RMP2009,Novoselov2005,Meng2010,Raghu2008}.
One of the most intriguing properties of graphene is that its chemical
potential can be tuned through an electric field effect,
and hence it is possible to change the type of
carriers, electrons, or holes, opening the doors for carbon based
electronics\cite{Novoselov2005,VHS,Ma2010}.
Doped graphene has a finite density of
state at the chemical potential, which, in combination with pronounced
antiferromagnetic (AFM) spin fluctuations close to half
filling\cite{Peres2004,Paiva2005}, may lead to an unconventional
superconductivity. Experimentally, superconducting
(SC) states in graphene have been realized by the proximity effect through
contact with SC electrodes\cite{Heersche2007},
which indicates that Cooper pairs can propagate coherently in graphene.
These facts raise the question as to whether it would be possible
to modify graphene to be an intrinsic superconductor.

Various theoretical attempts\cite{Uchoa2007,Jiang2008,carsten2008,schaffer2007,
Baskaran2002,Baskaran2010} have been made to understand the superconductivity
in graphene. Uchoa \emph{et al.}\cite{Uchoa2007} suggested that an
extended-$s$ (ES) SC phase may be realized at the mean-field
level due to the special structure of the honeycomb lattice.
On the other hand, in a weak-coupling functional
renormalization group study\cite{carsten2008}, Honerkamp found that with
a nearest-neighbor (NN) spin-spin interaction $J$, doping away from
half filling can lead to a $d+id$ SC state,
which is similar to the SC state on the triangular
lattice\cite{Kumar2003,Wang2004}. This $d+id$ SC state was also
found to be stable in a mean-field study\cite{schaffer2007} of
a phenomenological Hamiltonian\cite{Baskaran2002}.
Recent variational Monte Carlo simulations of the repulsive Hubbard model
provide further support for the $d+id$ SC state\cite{Baskaran2010}.

Although the results based on mean-field theory and other approximate methods
are encouraging, it is far from certain that there exists
a SC ground state in the physical parameter region of graphene.
It is well known that the low energy properties of graphene can be
described by the two-dimensional Hubbard model on a honeycomb
lattice~\cite{RMP2009}. In graphene, the on-site Hubbard repulsion is approximately
half the band width, and this places graphene in an intermediate-coupling
regime. Thus, it is questionable to approach the effect of electron
correlations in graphene from either a weak-coupling or strong-coupling
limit, as was done in many previous theoretical studies.
In view of the above-mentioned facts, we employ two
accurate numerical methods,
i.e., the determinant quantum Monte Carlo (DQMC)\cite{Blankenbecler1981}
and constrained path Monte Carlo (CPMC) methods\cite{Zhangcpmc,Huangcpmc},
to investigate possible electron correlation driven superconductivity
in graphene.

Our extensive numerical simulations show that close to half filling,
pairing with $d+id$ symmetry is dominant over pairing with ES
symmetry. However, the long-range part of the $d+id$ pairing correlation
tends to vanish in the thermodynamic limit, suggesting the absence of
electron correlation driven superconductivity in our studied model.
We also find that the NN interaction, either repulsive
or attractive, does not enhance the tendency to the $d+id$ or
ES superconductivity.

The structure of graphene can be described in terms of two interpenetrating
triangular sublattices, A and B, and its low-energy electronic and magnetic
properties can be well described by the extended Hubbard model
on a honeycomb lattice~\cite{RMP2009},
\begin{eqnarray}
H&=&-t\sum_{i\eta\sigma}a_{i\sigma}^{\dagger} b_{i+\eta\sigma}+ \mathrm{h.c.}%
+U\sum_{i}(n_{ai\uparrow}n_{ai\downarrow} + n_{bi\uparrow}n_{bi\downarrow})
\notag \\
&&+V\sum_{i\eta}n_{ai}n_{bi+\eta}+\mu\sum_{i\sigma}(n_{ai\sigma }+n_{bi\sigma }),
\end{eqnarray}
Here, $a_{i\sigma}$ ($a_{i\sigma}^{\dag}$) annihilates (creates) electrons
at site $\mathbf{R}_i$ with spin $\sigma$ ($\sigma$=$\uparrow,\downarrow$)
on sublattice A, $b_{i\sigma}$ ($b_{i\sigma}^{\dag}$) annihilates (creates)
electrons at the site $\mathbf{R}_i$ with spin $\sigma$
($\sigma$=$\uparrow,\downarrow$) on sublattice B,
$n_{ai\sigma}=a_{i\sigma}^{\dagger}a_{i\sigma}$ and
$n_{bi\sigma}=b_{i\sigma}^{\dagger}b_{i\sigma}$.
$t\approx 2.5eV$ is the NN hopping integral and
$\mu$ the chemical potential. $U\approx 6 eV$ and $V$ denote the on-site
Hubbard interaction and NN interaction, respectively.
We have mainly used $U=3|t|$ and $V=0$ in this Rapid Communication,
except as explicitly noted otherwise.

\begin{figure}[tbp]
\includegraphics[scale=0.64]{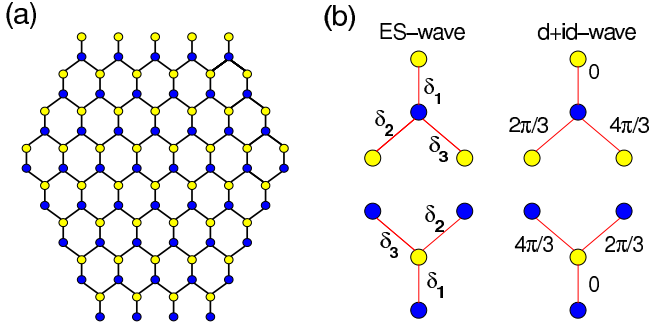}
\caption{(Color online) (a) Sketch of graphene with double-48 sites,
 (b) Phases of the $d+id$ and $ES$ pairing symmetries on the honeycomb
lattice. For clarity, the phase for the $ES$ symmetry is omitted, which
equals $0$ for all $\delta_{l}$.}
\label{grapair}
\end{figure}

Our numerical calculations are performed on lattices of double-48,
double-75, double-108, and double-147 sites with periodic boundary
conditions. The double-48 lattice is sketched in Fig.~\ref{grapair}(a),
where blue circles and yellow circles indicate the A and B sublattices,
respectively. The system is simulated using DQMC at finite temperature and
CPMC at zero temperature. The basic strategy of DQMC is to express
the partition function as a high-dimensional integral over a set of
random auxiliary fields. The integral is then accomplished by
Monte Carlo techniques. In the CPMC method, the ground-state wave
function is projected from an initial wave function by a branching
random walk in an overcomplete space of constrained Slater
determinants, which have positive overlaps with a known trial wave
function. Extensive benchmark calculations showed that the systematic
error induced by constraint is within a few percent and the ground-state
observables are insensitive to the choice of trial wave
function~\cite{Zhangcpmc}.
In our CPMC simulations, we employ closed-shell electron fillings and use
the corresponding free-electron wave function as the trial wave function.

As magnetic excitation might play an important role in the SC mechanism
of electronic correlated systems, we first study the magnetic
correlations in graphene. Specifically, we compute the NN
spin correlation $S_{<i,j>}^{zz}=\langle S_{i}^{z}\cdot S_{j}^{z}\rangle$
and the spin structure factor $S(\textbf{q})$, which is defined as,
\begin{equation}
S(\mathbf{q}) =\frac{1}{N_{s}} \sum_{d,d^{\prime }=a,b} \sum_{i,j}
e^{iq\cdot(i_{d}-j_{d^{\prime }})} \langle\text{m}_{i_{d}} \cdot \text{%
m}_{j_{d^{\prime }}}\rangle,
\end{equation}
where $m_{i_{a}}=a^{\dag}_{i\uparrow}a_{i\uparrow}-a^{\dag}_{i\downarrow}
a_{i\downarrow}$ and $m_{i_{b}}=b^{\dag}_{i\uparrow}b_{i\uparrow}-b^{\dag}_{i
\downarrow}b_{i\downarrow}$. $N_{s}$ denotes the number of sublattice sites.

To investigate the SC property of graphene, we compute the pairing
susceptibility,
\begin{equation}
P_{\alpha}=\frac{1}{N_s}\sum_{i,j}\int_{0}^{\beta }d\tau \langle \Delta
_{\alpha }^{\dagger }(i,\tau)\Delta _{\alpha }^{\phantom{\dagger}%
}(j,0)\rangle,
\end{equation}
and the pairing correlation,
\begin{eqnarray}
C_{\alpha }({\bf{r=R_{i}-R_{j}}})=\langle \Delta _{\alpha }^{\dagger }
(i)\Delta _{\alpha }^{\phantom{\dagger}}(j)\rangle ,
\end{eqnarray}
where $\alpha$ stands for the pairing symmetry.
Due to the constraint of the on-site Hubbard interaction in
Eq.~(1), pairing between two sublattices is favored and the corresponding
order parameter $\Delta _{\alpha }^{\dagger }(i)$\ is defined as
\begin{eqnarray}
\Delta _{\alpha }^{\dagger }(i)\ =\sum_{l}f_{\alpha}^{\dagger}
({\bf\delta}_{l})(a_{{i}\uparrow }b_{{i+{\bf\delta}_{l}}\downarrow }-
a_{{i}\downarrow}b_{{i+{\bf\delta}_{l}}\uparrow })^{\dagger},
\end{eqnarray}
with $f_{\alpha}(\bf{\delta}_{l})$ being the form factor of pairing function.
Here, the vectors $\bf{\delta_{l}}$ $(l=1,2,3)$ denote the NN
intersublattice connections, as sketched in Fig.~\ref{grapair}(b).
Considering that the pairing symmetry of graphene is governed by the
$D6$ point group, two form factors for NN pairing described by the $A_1$
and $E_2$ irreducible representations of the $D6$ point group are given
by~\cite{Jiang2008},
\begin{eqnarray}
\text{$ES$-wave} &\text{:}&\ f_{ES}({\bf\delta}_{l})=1,~l=1,2,3 \\
\text{$d+id$-wave} &\text{:}&\ f_{d+id}({\bf\delta}_{l})=e^{i(l-1)
\frac{2\pi }{3}},~l=1,2,3
\end{eqnarray}



\begin{figure}[tbp]
\includegraphics[scale=0.4]{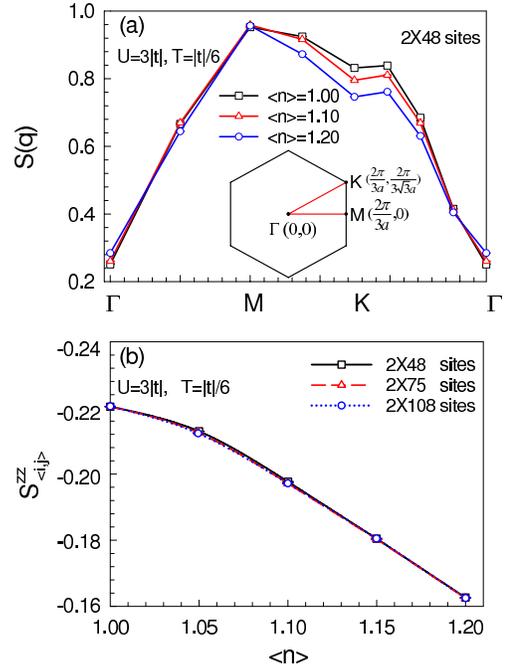}
\caption{(Color online) (a) Spin structure factor $S(\textbf{q})$ on
the double-48 lattice along the high symmetry lines in the first
Brillouin zone (BZ) at electron fillings $<n>=1.00, 1.10$, and $1.20$.
(b) NN spin correlation $S_{<i,j>}^{zz}$ versus $<n>$ on different lattices.
The inset of (a) shows the first BZ, with $a$ denoting
the distance between NN lattice sites. The results are presented for
temperature $T$=$\mid$t$\mid$/6.}
\label{corrspin}
\end{figure}

In Fig. \ref{corrspin}, we present the spin structure factor
$S(\textbf{q})$ and the NN spin correlation $S_{<i,j>}^{zz}$
at different electron fillings for temperature $T$=$\mid$$t$$\mid$/6.
A broad peak between $M$ and $K$ in Fig.~\ref{corrspin}(a) and a negative NN
spin correlation in Fig.~\ref{corrspin}(b) indicate the existence of
AFM spin correlation in graphene close to half filling.
From Fig.~\ref{corrspin}(a), one can notice that even at half-filling,
the peak is rather weak, suggesting that magnetic ordering is strongly
frustrated due to the structure of the honeycomb lattice.
Moreover, Fig. \ref{corrspin} shows that as the electron filling $<n>$
increases from half filling, $S(\textbf{q})$ is reduced in the region around
the $K$ point and $S_{<i,j>}^{zz}$ becomes less negative, which indicates that
the AFM spin correlation is suppressed when the system is doped away from
half filling. As it is expected that fermion systems with strong on-site
repulsion may exhibit superconductivity induced by AFM spin fluctuations,
and from the behavior of magnetic correlation shown in Fig.~\ref{corrspin},
it seems that the electron correlation driven superconductivity is possible
through a similar mechanism in graphene. In the following, we discuss the
behavior of pairing susceptibility and pairing correlation in the low
doping region.

\begin{figure}[tbp]
\includegraphics[scale=0.5]{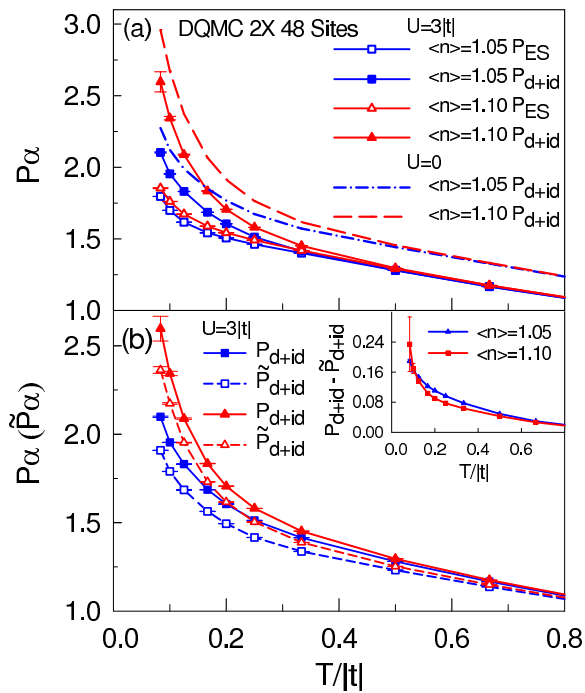}
\caption{(Color online) (a) Pairing susceptibility $P_{\alpha}$ as a function
of temperature $T$ for different pairing symmetries and different fillings,
(b) $\tilde{P}_{d+id}$ and $P_{d+id}$ as a function of temperature
at $<n>=1.05$ and $<n>=1.10$. Dashed and dashed-dotted lines in (a)
represent the $d+id$ pairing susceptibility at $U=0$.
The inset of (b) shows the effective pairing interaction
$P_{d+id}-\tilde{P}_{d+id}$ as a function of temperature.}
\label{Fig:DS}
\end{figure}

Figure~\ref{Fig:DS} shows the temperature dependence of pairing
susceptibilities for different pairing symmetries and electron fillings
on the double-48 lattice.
From Fig.~\ref{Fig:DS}(a), it is clear to see that within the filling range
investigated, the pairing susceptibilities for both $d+id$ and $ES$ pairing
symmetries increase as the temperature is lowered. Most remarkable is that
$P_{d+id}$ increases much faster than $P_{ES}$ at low temperatures.
This demonstrates that the $d+id$ pairing symmetry is dominant over the
$ES$ pairing symmetry in the low doping region. In the whole temperature
regime, one can observe that the value of $P_{d+id}$ at $U=3|t|$ is smaller
than the corresponding noninteracting one ($U=0$), as displayed in
Fig.~\ref{Fig:DS}(a), which reflects the fact that the reduction of
quasiparticle weight (self-energy effect) due to electron correlation
plays a negative role in enhancing the pairing susceptibility.

In order to extract the effective pairing interaction in
different pairing channels, the bubble contribution $\widetilde{P}
_{\alpha }(i,j)$ is also evaluated, which is achieved by replacing $\langle
a_{{i}\downarrow }^{\dag }b_{{j}\uparrow }a_{i+{\bf\delta}_{l}\downarrow}^{\dag}
b_{j+{\bf\delta}_{l'}\uparrow}\rangle $ with $\langle a_{{i}\downarrow }^{\dag
}b_{{j}\uparrow }\rangle \langle a_{i+{\bf\delta}_{l}\downarrow }^{\dag }
b_{j+{\bf\delta}_{l'}\uparrow }\rangle $ in Eq.~(3). In Fig. \ref{Fig:DS} (b), we
plot both $P_{d+id}$ and $\widetilde{P}_{d+id}$ for comparison.
It is apparent that $\widetilde{P}_{d+id}$ shows a very similar temperature
dependence to that of $P_{d+id}$.
The effective pairing interaction, which can be
estimated by the difference between $P_{d+id}$ and $\widetilde{P}_{d+id}$,
is found to take a positive value and to increase with lowering temperature,
as clearly shown in the inset figure.
The positive effective pairing interaction indicates that there actually
exists attraction for the $d+id$ pairing.

Based on the DQMC results for the pairing susceptibility, one expects
that there may exist the $d+id$ SC state in the low-temperature region,
which is manifested by a divergence of $P_{d+id}$ at a certain temperature.
Unfortunately, it is not clear whether the pairing susceptibility keeps
growing at low temperatures since the sign problem prevents simulation in the
low-temperature regime. In order to shed light on the critical issue as to
whether there exists a long-range off-diagonal $d+id$ SC order in the ground
state, we now turn to discuss the results obtained from the CPMC method.

\begin{figure}[tbp]
\epsfig{file=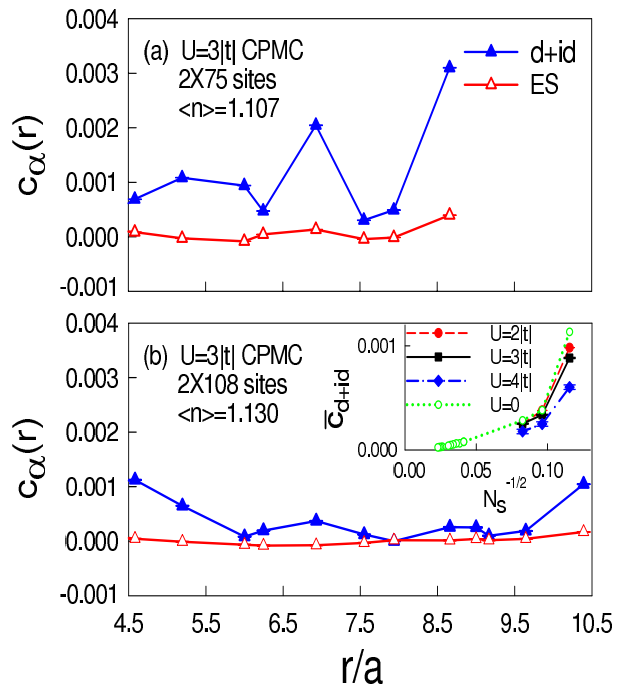,height=9cm,width=8cm,angle=0}
\caption{(Color online) Pairing correlation $C_{\alpha}$ as a function of
distance for the $d+id$ and $ES$ pairing symmetries on the double-75
lattice with $<n>=1.107$ (a) and double-108 lattice with $<n>=1.13$ (b).
The inset of (b) shows the average of the long-range $d+id$
pairing correlation $\overline{C}_{d+id}$ vs $\frac{1}{\sqrt{N_s}}$
for different $U's$ at a filling $<n>\approx 1.1$.}
\label{Fig:PDr}
\end{figure}

In Fig.~\ref{Fig:PDr}, we compare the long-range part of pairing
correlations for the $d+id$ and $ES$ pairing symmetries on the double-75
and double-108 lattices, with a filling of approximately $<n>=1.1$.
One can readily see that $C_{d+id}(r)$ is
larger than $C_{ES}(r)$ for all long-range distances between electron
pairs. Similar behavior is also observed on the double-147 lattice
with $<n>=1.095$ (not shown here). This re-enforces our finding that
the $d+id$ pairing symmetry dominates the $ES$ pairing symmetry
in the low doping region.

To gain insight into the behavior of the $d+id$ pairing correlation
in the thermodynamic limit, we examine the evolution of $C_{d+id}$
with increasing lattice size.
In the inset of Fig. \ref{Fig:PDr}(b), the average of the
long-range $d+id$ pairing correlation, $\overline{C}_{d+id}=
\frac{1}{\sqrt{N'}}\sum_{r>4a}C_{d+id}(r)$, where $N'$ is the number
of electron pairs with $r>4a$, is plotted as a function of
$\frac{1}{\sqrt{N_s}}$ for $U=0, 2|t|, 3|t|$, and $4|t|$.
We observe that $\overline{C}_{d+id}$ decreases as the lattice size
increases, and shows a clear tendency to vanish in the thermodynamic
limit ($\frac{1}{\sqrt{N_s}}\rightarrow 0$) at $U=0$.
This result, together with a decrease of $\overline{C}_{d+id}$ with
increasing $U$, suggests the absence of long-range
$d+id$ SC order in the parameter regime investigated.
Our finding is in agreement with the functional
renormalization group study\cite{carsten2008}, where the SC instability
does not occur in the doped Hubbard model. A similar decrease
of $\overline{C}_{d+id}$ with increasing lattice size was also
observed in the variational Monte Carlo calculations~\cite{Baskaran2010}.
Thus, although AFM fluctuations can mediate a NN singlet pair,
quantum fluctuations beyond the mean-field level shall destroy
the phase coherence between electron pairs.

\begin{figure}[tbp]
\includegraphics[scale=0.5]{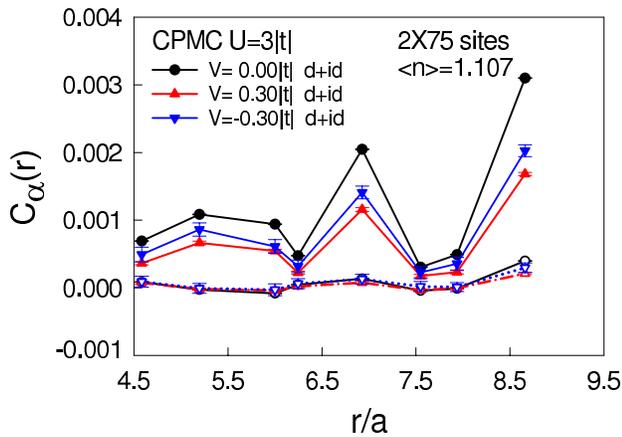}
\caption{(Color online) Pairing correlation as a function of distance
for the $d+id$ (filled symbols) and $ES$ (open symbols)
pairing symmetries on the double-75 lattice
with $<n>=1.107$. The value of NN interaction $V$ is indicated by the
shape of the symbol, which is also applied for the $ES$ symmetry.}
\label{Fig:PdV}
\end{figure}
In graphene, the long-range interaction may also play a significant
role on its physical properties, especially in the low doping region,
where the long-range Coulomb interaction is not effectively screened
because of a small density of state at the Fermi energy.
We have studied the effect of NN interaction on the pairing correlation.
In Fig. \ref{Fig:PdV}, the pairing correlations for both $d+id$ and
$ES$ pairing symmetries are displayed as a function of distance
on the double-75 lattice with different NN interaction $V's$.
Here, we consider both repulsive and attractive NN interactions.
We notice that the $ES$ pairing correlation is hardly affected by
the NN interaction, whereas the $d+id$ pairing correlation is suppressed
by either a repulsive or attractive NN interaction.
Our results are contrary to the finding of
Uchoa \emph{et al.}\cite{Uchoa2007}, where the NN attraction can
stabilize the $ES$ SC state at the mean-field level.
In addition, the inclusion of NN interaction does not
enhance the tendency to the $d+id$ SC state.
Therefore, we can conclude that within the extended Hubbard model,
there seems to be an absence of SC order in the ground state.


In summary, we have studied the behavior of pairing correlation
within the extended Hubbard model on a honeycomb lattice by using
quantum Monte Carlo simulations. The results obtained from both
DQMC and CPMC show that close to half filling, pairing with $d+id$
symmetry dominates pairing with ES symmetry, which is consistent
with previous mean-field and functional renormalization group studies.
This provides strong evidence that pairing with
different spatial phases is favored for the AFM fluctuation mediated
pairing interaction, which is similar to the case on the
triangular lattice~\cite{Kumar2003,Wang2004}.
However, the tendency of the $d+id$ pairing
correlation to vanish in the thermodynamic limit suggests that
electron correlation in graphene is not strong enough to produce
an intrinsic superconductivity.  In an induced SC state by the
proximity effect through a connection to another superconductor,
the dominant $d+id$ pairing could be manifested in the unique
properties of the SC gap function and the Andreev conductance
spectra~\cite{Jiang2008}.

This work is partially supported by HKSAR RGC Project No. CUHK 402310.
Z.B.H was supported by NSFC Grant No. 10974047.

\end{document}